\documentclass[12pt]{article}
\usepackage{amsmath}
\usepackage{amsfonts}
\usepackage{amssymb}
\usepackage{latexsym}
\usepackage{graphicx}
\usepackage[english]{babel}
\input epsf.sty

\begin{document}

\topmargin 0pt
\oddsidemargin 0mm



\begin{flushright}

\hfill{USTC-ICTS-07-25}\\

\end{flushright}
\vspace{4mm}

\begin{center}

{\Large \bf General Single Field Inflation with\\ Large Positive Non-Gaussianity}

\vspace{8mm}

{\large Miao Li$^{1,2}$, Tower Wang$^{2,1}$, Yi Wang$^{2,1}$}

\vspace{4mm}

{\em
$^{1}$ Interdisciplinary Center of Theoretical Studies, USTC,\\
Hefei, Anhui 230026, P.R.China\\
$^{2}$ Institute of Theoretical Physics, CAS,\\
Beijing 100080, P.R.China}

\end{center}

\vspace{6mm}

Recent analysis of the WMAP three year data suggests
$f_{NL}^{local}\simeq86.8$ in the WMAP convention. It is necessary
to make sure whether general single field inflation can produce a
large positive $f_{NL}$ before turning to other scenarios. We give
some examples to generate a large positive $f_{NL}^{equil}$ in
general single field inflation. Our models are different from ghost
inflation. Due to the appearance of non-conventional kinetic terms,
$f_{NL}^{equil}\gg1$ can be realized in single field inflation.

\newpage


\section{Introduction}
As modern cosmological observations become more and more precise,
study of the non-Gaussianity of CMB temperature fluctuations has
become a more and more pressing issue in recent years. In the WMAP
convention \cite{Komatsu:2001rj,Spergel:2006hy}, the primordial
non-Gaussianity is parameterized by $f_{NL}$ assuming the ansatz
\begin{equation}\label{wmap}
\zeta=\zeta_L+\frac{3}{5}f_{NL}\zeta_L^2,
\end{equation}
where $\zeta$ is the scalar perturbation and $\zeta_L$ is its linear
Gaussian part\footnote{Note that Maldacena's convention is also
popularly used in the literature \cite{Maldacena:2002vr,Chen:2006nt},
which assumes $\zeta=\zeta_L-\frac{3}{5}f_{NL}\zeta_L^2$ instead of
(\ref{wmap}). That convention is different in sign of $f_{NL}$ from
the WMAP convention. Please refer to
\cite{Komatsu:2002db,Komatsu:kitpc} for clarification. In this
paper, we will use the WMAP convention.}. However, this ansatz only
corresponds to a restricted shape of non-Gaussianity. Theoretically,
during inflation, the non-Gaussianity is usually produced in different shapes, and the
estimator $f_{NL}$ can be defined  in terms of the 3-point function
$\langle\zeta_{k_1}\zeta_{k_2}\zeta_{k_3}\rangle$. Two limits of $f_{NL}$ are of
most interest. One is the local, squeezed limit $k_1\ll k_2\simeq
k_3$, for which we will use the notation $f_{NL}^{local}$. The other
is the non-local, equilateral limit $k_1\simeq k_2\simeq k_3$, for
which the notation will be $f_{NL}^{equil}$.

In a recent work \cite{Yadav:2007yy}, utilizing the fast estimator
of primordial non-Gaussianity \cite{Yadav:2007ny}, Yadav and Wandelt
claimed that data from two channels of WMAP3 reject
$f_{NL}^{local}=0$ at the $2.89\sigma$ level, or $99.6\%$
significance. They also showed that $26.91<f_{NL}^{local}<146.71$ at
$95\%$ C.L., with a central value of $f_{NL}^{local}=86.8$. If this
result is confirmed by future observations, it will have a great
impact on our study of the early universe, because a large class of
inflation models will be ruled out. For example, the simplest model
of inflation is slow rolling, driven by a single scalar field.
Ignoring the non-Gaussianity, previous observational data fit well
with single-field slow-roll inflation
\cite{Spergel:2006hy,Alabidi:2006qa}. On the other hand, in the
conventional single-field slow-roll inflation, it has been found for
both local and equilateral forms that $|f_{NL}|<1$, which is too
small to detect in the near future
\cite{Maldacena:2002vr,Acquaviva:2002ud}. Therefore, the
confirmation of a large $f_{NL}^{local}$ observationally will
exclude almost all models of single-field slow-roll inflation.

Confronted with the evidence for $f_{NL}^{local}\gg1$, is
single-field slow-roll inflation dying? Not necessarily. There are
at least two ways to save it. First, a large non-Gaussianity may
arise in stochastic inflation due to non-linear effects between
inflaton and metric perturbations \cite{Gangui:1993tt}. Second, the
curvaton mechanism provides an elegant way to produce a large
positive $f_{NL}^{local}$ \cite{Lyth:2002my}.

Although there is no evidence for $|f_{NL}^{equil}|$ to be large at
present, several inflation models with $|f_{NL}^{equil}|\gg1$ have
already appeared in the past few years. In single-field slow-roll
inflation, this is realized by the introduction of non-canonical
kinetic terms, such as $k$-inflation
\cite{ArmendarizPicon:1999rj,Garriga:1999vw}, ghost inflation
\cite{ArkaniHamed:2003uz} DBI inflation
\cite{Silverstein:2003hf,Alishahiha:2004eh,Chen:2004gc,Chen:2005ad},
and some other mechanisms \cite{berera}. Translated into the WMAP
convention, most of the models predict $f_{NL}^{equil}\ll-1$ and a
small $f_{NL}^{local}$. It is remarkable that the value of
$f_{NL}^{local}$ favored by \cite{Yadav:2007yy} is of positive sign
in the WMAP convention. For $f_{NL}^{equil}$, the present constraint
is not stringent enough to make a conclusion. So one expects that
there are three possibilities in the future:
\begin{enumerate}
\item $f_{NL}^{local}\gg1$, $-1<f_{NL}^{equil}<1$. This can be
explained by the conventional single-field slow-roll inflation +
curvaton mechanism.
\item $f_{NL}^{local}\gg1$, $f_{NL}^{equil}<-1$. This corresponds
to DBI inflation (or most known $k$-inflation/ghost inflation) +
curvaton mechanism.
\item $f_{NL}^{local}\gg1$, $f_{NL}^{equil}>1$. This is more challenging to
explain.
\end{enumerate}

For both the second and the third possibility, one should keep in mind
that additional fine-tuning is needed in order that both the curvaton and
the inflaton produce perturbations of comparable magnitude.

The purpose of this paper is to search for $k$-inflation models with
$f_{NL}^{equil}\gg1$. If this type of models are constructed, one
can combine them with the curvaton mechanism to meet the
challenge posed by the third possibility.

Indeed, starting from the action (\ref{action}), we have found
several examples of general single field inflation in which
$f_{NL}^{equil}\gg1$. As will be shown in section \ref{run}, even if
the inflation is driven by the potential of inflaton, a large
positive $f_{NL}^{equil}$ can be generated by non-conventional
kinetic terms. In some other models, the non-conventional kinetic
terms not only give rise to the desired $f_{NL}^{equil}$, but also
drive the inflation. Examples for this type of models are
constructed in power-law $k$-inflation. In all our models, typically
the desired non-Gaussianity stems from high order terms in $X$, with
$X$ defined in (\ref{X}).

The paper is organized as follows. In section \ref{review}, we
review briefly the general single field inflation. In section
\ref{nogo}, we prove a no-go theorem for the $p(X)$ models. We show
that if the matter Lagrangian depends only on $X$, but not on $\phi$
directly, then one can not obtain a large and positive
$f^{equil}_{NL}$. In section \ref{run}, we construct generalized
slow roll inflation models with $f^{equil}_{NL}\gg 1$. In section
\ref{pl}, we construct power-law inflation models with
$f^{equil}_{NL}\gg 1$. We conclude in section \ref{conclusion}.

\section{A Brief Review of General Single Field Inflation}\label{review}
In this section, we give a brief review on the general single field
inflation. For further details, please refer to
\cite{Chen:2006nt,ArmendarizPicon:1999rj,Garriga:1999vw,Seery:2005wm}.
We consider the action \cite{ArmendarizPicon:1999rj}
\begin{equation}\label{action}
S=\int d^4x\sqrt{-g}[\frac{1}{16\pi G}R+p(\phi,X)],
\end{equation}
where
\begin{equation}\label{X}
X=-\frac{1}{2}g^{\mu\nu}\partial_{\mu}\phi\partial_{\nu}\phi,
\end{equation}
and the signature of metric is $(-1,1,1,1)$. We henceforth set the
reduced Planck mass $M_{pl}=(8\pi G)^{-\frac{1}{2}}$ to unity.

In terms of the pressure $p(\phi,X)$ and its derivatives with
respect to $X$ (denoted by $p_{,X}$ \emph{etc.}), we can write down
the energy density
\begin{equation}\label{rho}
\rho(\phi,X)=2Xp_{,X}-p
\end{equation}
of the inflaton as well as the speed of sound
\begin{equation}\label{soundspd}
c_s^2=
\frac{p_{,X}}{p_{,X}+2Xp_{,XX}}.
\end{equation}
The Friedmann equation and the continuity equation are given by
\begin{eqnarray}\label{Friedmann}
\nonumber 3H^2&=&\rho,\\
\dot{\rho}&=&-3H(\rho+p).
\end{eqnarray}
It proves useful to define two quantities
\begin{eqnarray}\label{siglam}
\nonumber \Sigma&=&Xp_{,X}+2X^2p_{,XX},\\
\lambda&=&X^2p_{,XX}+\frac{2}{3}X^3p_{,XXX},
\end{eqnarray}
and some slow-variation parameters
\begin{eqnarray}\label{slowvar}
\nonumber &&\epsilon=-\frac{\dot{H}}{H^2}=\frac{3Xp_{,X}}{2Xp_{,X}-p},\\
&&\eta=\frac{\dot{\epsilon}}{H\epsilon},~~~~s=\frac{\dot{c_s}}{Hc_s},~~~~l=\frac{\dot{\lambda}}{H\lambda}
\end{eqnarray}
following \cite{Chen:2006nt,Seery:2005wm}. We make note that one of
the ``slow-variation'' parameters $\eta$ here is different from one
of the ``slow-roll'' parameters frequently used in ordinary
slow-roll inflation. On this point please see \cite{Seery:2005wm}
for clarification. Throughout our discussion, we will be interested
only in the the slow-variation case $\epsilon,\eta,s,l\ll1$ with
$\dot{H}\leq0$, $\epsilon\geq0$.

According to \cite{Garriga:1999vw}, to the leading order, the power
spectra for scalar and tensor perturbations are
\begin{eqnarray}\label{sptrm}
\nonumber P_k^{\zeta}&=&\frac{H^2}{8\pi^2c_s\epsilon},\\
P_k^{h}&=&\frac{2H^2}{\pi^2},
\end{eqnarray}
which lead to the tensor-to-scalar ratio
\begin{equation}\label{r}
r=\frac{P_k^{h}}{P_k^{\zeta}}=16c_s\epsilon.
\end{equation}
While their spectral indices are
\begin{eqnarray}\label{nsnt}
\nonumber n_s-1&=&-2\epsilon-\eta-s,\\
n_T&=&-2\epsilon.
\end{eqnarray}

In accordance with the WMAP convention (\ref{wmap}), the
non-Gaussianity parameter $f_{NL}$ in the equilateral triangle limit
is \cite{Chen:2006nt}
\begin{equation}\label{fNL}
f_{NL}^{equil}=-\frac{10}{81}\frac{\lambda}{\Sigma}+\left(\frac{5}{81}-\frac{35}{108}\right)\left(\frac{1}{c_s^2}-1\right)+\frac{5(3-2\gamma)}{81}\frac{l\lambda}{\Sigma}+\mathcal{O}\left(\frac{\epsilon}{c_s^2},\frac{\epsilon\lambda}{\Sigma}\right)+\mathcal{O}(\epsilon).
\end{equation}
Here $\gamma$ is the Euler-Mascheroni constant, which is denoted by
$\mathbf{c}_1$ in \cite{Chen:2006nt}. Numerically
$\gamma\simeq0.577$. Please always keep in mind that we follow the
WMAP convention hence the $f_{NL}^{equil}$ here is opposite in sign
with respect to that in \cite{Maldacena:2002vr,Chen:2006nt}. From
(\ref{fNL}) it is clear that in order to get a large positive
$f_{NL}^{equil}$, we should have $-\frac{\lambda}{\Sigma}\gg1$.
Neglecting the sub-leading terms, one can naively estimate
$-\frac{10}{81}\frac{\lambda}{\Sigma}$ with $f_{NL}^{equil}$, thus
we have $-\frac{\lambda}{\Sigma}\sim\mathcal{O}(10^3)$ in order to
get $f_{NL}^{equil}\sim 100$. This observation will be useful in our
model reconstruction below.

Note that in the case $-\frac{\lambda}{\Sigma}\gg1$, the ${\cal
O}(\epsilon\lambda/\Sigma)$ correction also gives a ${\cal O}(1)$
contribution. So to be more precise, we have
\begin{eqnarray}\label{fnlprecise}
f_{NL}^{equil}&=&-\frac{10}{81}\frac{\lambda}{\Sigma}+\left(\frac{5}{81}-\frac{35}{108}\right)\left(\frac{1}{c_s^2}-1\right)+\frac{5(3-2\gamma)}{81}\frac{l\lambda}{\Sigma}\nonumber\\
&&-\frac{5}{243}\frac{\lambda}{\Sigma}\left\{\left(2\epsilon+\eta+s\right)\left(-3\gamma-48+\frac{252}{2}\ln
\frac{3}{2}\right)-\eta\left(6\gamma-\frac{33}{2}\right)+s(3\gamma-12)\right\}\nonumber\\
&&+{\cal O}\left(\epsilon,~ \frac{\epsilon}{c_s^2},~
\frac{\epsilon^2\lambda}{\Sigma}\right)~,
\end{eqnarray}
where we have performed the integral in $R(k_1, k_2, k_3)$
\cite{Chen:2006nt} in the equilateral limit $k_1=k_2=k_3$. The
details of the integration is given in Appendix \ref{CalRk}. In
(\ref{fnlprecise}), we neglect $\mathcal{O}(\frac{\epsilon}{c_s^2})$
terms by concentrating on the parameter space $c_s^2\gg\epsilon$. If
one would like to consider models with a smaller $c_s^2$, those
terms should be taken into account.

\section{A No-Go Result for $p(X)$ Model}\label{nogo}
In the simplest case, Lagrangian $p$ takes the form $p=p(X)$
independent of $\phi$. This model mimics a de Sitter space, in which
the cosmological perturbations are ``ill-defined''
\cite{ArmendarizPicon:1999rj}. This model cannot give large positive
$f_{NL}^{equil}$, as we now show by applying the general
results in section \ref{review} to it.

Making use of (\ref{rho}-\ref{slowvar}), it is not hard to check for
$p=p(X)$ that
\begin{eqnarray}\label{slowvardS}
\nonumber s&=&\frac{\dot{X}}{2HX}\left(\frac{X^2p_{,XX}}{Xp_{,X}}-\frac{3X^2p_{,XX}+2X^3p_{,XXX}}{Xp_{,X}+2X^2p_{,XX}}\right)\\
\nonumber &=&-\frac{3Xp_{,X}}{Xp_{,X}+2X^2p_{,XX}}\left(\frac{X^2p_{,XX}}{Xp_{,X}}-\frac{3X^2p_{,XX}+2X^3p_{,XXX}}{Xp_{,X}+2X^2p_{,XX}}\right)\\
&=&9c_s^2\left(\frac{\lambda}{\Sigma}+\frac{1}{6}\right)-\frac{3}{2}.
\end{eqnarray}
It is obvious from (\ref{fNL}) that a large positive $f_{NL}^{equil}$
implies $\frac{\lambda}{\Sigma}\ll-1$. This requirement leads to
$s<-1$ in (\ref{slowvardS}), violating the slow-variation
condition.

So a large positive $f_{NL}^{equil}$ cannot appear in this simple
model.

For a more general Lagrangian $p=p(\phi,X)$, terms like
$(2Xp_{,X\phi}-p_{,\phi})\dot{\phi}$ will show up in the continuity
equation (\ref{Friedmann}), hence the above relation does not hold
anymore, and one should study case by case. For some special forms
of $p(\phi,X)$, one may get a relation similar to (\ref{slowvardS})
and a no-go theorem likewise. For other cases, as we will
investigate in the following sections, such a no-go theorem does not
exist and we can construct slow-variation inflation models with
$\frac{\lambda}{\Sigma}\ll-1$.

\section{Reconstruction of the Generalized Slow Roll Inflation}\label{run}
In this section, we investigate the non-Gaussianity estimator
$f_{NL}^{equil}$ of the generalized slow roll inflation. We will
show that a large and positive $f_{NL}^{equil}$ can be obtained in
relatively simple models of this class. By generalized slow roll
inflation models, we mean that inflation is still driven by the
potential energy of inflaton, while the inflaton has generalized
kinetic terms, which can generate large non-Gaussianities. For this
purpose, we study the Lagrangian
\begin{equation}
  p(\phi, X)=g(\phi) f(X)-V(\phi)~.
\end{equation}
Using (\ref{rho}), the energy density can be written as
\begin{equation}
  \rho=2gXf_{,X}-gf+V~.
\end{equation}
For our purpose, we look for solutions with $|2gXf_{,X}| \ll V$ and $|gf|
\ll V$. The validity of this ansatz will be checked later in this
section. Then to the leading order approximation, we have
\begin{equation}
  \rho \simeq V~.
\end{equation}
The equation of motion of $\phi$ can be written as
\begin{equation}\label{varphieom}
  \partial_t(gf_{,X}\dot\phi)+3H
  gf_{,X}\dot\phi-g_{,\phi}f+V_{,\phi}=0~.
\end{equation}
For simplicity, we study solutions with $\ddot\phi\simeq 0$. This is
the direct generalization of the $\frac{1}{2}m^2\phi^2$ model with
the standard kinetic term. We expect this simplification does not
lose much generality for two reasons. Firstly, for slow roll
inflation, $\ddot\phi/(H\dot\phi)\ll 1$, so they commonly behave the
way we assume. Secondly, we are mainly interested in the
non-Gaussianity, which is generated mainly by $f(X)$. As we will
show, we can choose $g$ and $V$ so that the assumption
$\ddot\phi\simeq 0$ does not lose generality for $f(X)$.

After this approximation, and using the Friedmann equation,
(\ref{varphieom}) takes the form
\begin{equation}\label{li}
  g_{,\phi}\left(f-2X f_{,X}-({\rm sgn}\dot\phi)\sqrt{6VX}f_{,X}\frac{g}{g_{,\phi}}
  \right)=V_{,\phi}~,
\end{equation}
where $({\rm sgn}\dot\phi)$ denotes the sign of $\dot\phi$, which
comes from the square root $\dot\phi=({\rm sgn}\dot\phi)\sqrt{2X}$.
We demand the equation (\ref{li}) to boil down to an equation of only $X$, so
\begin{equation}\label{propli}
  \frac{g_{,\phi}}{g}\propto \sqrt{V}~,~~~ g_{,\phi} \propto
  V_{,\phi}~.
\end{equation}
The solution of (\ref{propli}) takes the form
\begin{equation}\label{soluli}
  g=-\frac{1}{\cosh^2(\alpha\phi)}~,~~~
  V=\frac{\alpha^2\beta^2}{3} \tanh^2(\alpha\phi)~,
\end{equation}
where $\alpha$ and $\beta$ are constants. Without losing generality,
we set $\alpha>0$, $\beta>0$, and $\phi>0$. In this case, $\phi$
rolls backwards, so $\dot\phi<0$.

Inserting the solution (\ref{soluli}) into (\ref{li}), we have the
equation for $X$
\begin{equation}\label{Xsolli}
  f-2Xf_{,X}-\beta\sqrt{\frac{X}{2}}f_{,X}=\frac{\alpha^2\beta^2}{3}~.
\end{equation}
We can take this equation either as a differential equation, which
is valid for all $X$, or as an algebraic equation, which is valid for
some certain $X$. It can be shown that the former possibility leads
to
\begin{equation}
 f=\frac{\alpha^2\beta^2}{3}+C\left(\frac{\beta}{\sqrt{2}}+2\sqrt{X}\right)~,
\end{equation}
where $C$ is a constant. In this case, $c_s \rightarrow \infty$ and
$\lambda/\Sigma\rightarrow 0/0$, so the next to leading order
contribution in the slow roll approximation must be taken into
consideration. In the remainder of this section, we consider the
latter possibility, and treat (\ref{Xsolli}) as an algebraic
equation.

Now let us verify the slow roll conditions. Compare (\ref{Xsolli})
with the energy density, one can find that if $f-2Xf_{,X}$ and
$\beta\sqrt{\frac{X}{2}}f_{,X}$ do not cancel at leading order of
slow roll approximation, then the condition $\alpha\phi\gg 1$ leads
to the slow roll condition $\epsilon\ll 1$. This condition can be
satisfied by imposing proper initial conditions.

It can be shown that
\begin{equation}
  \eta\simeq l\simeq\frac{6\sqrt{2X}}{\beta}.
\end{equation}
So the condition $\eta,l\ll 1$ can be satisfied by requiring that
$\beta$ is large enough.

The slow roll condition for $s$ is automatically satisfied, because
we have assumed $\dot X \propto \ddot\phi\simeq 0$, which is
verified in (\ref{Xsolli}) where $X$ is a constant.

To solve (\ref{Xsolli}), we need to give an explicit expression for
$f(X)$. As an illustration, we consider the simplest polynomial case
\begin{equation}
  f=c_1 X + c_2 X^2~,
\end{equation}
the calculation can be generalized to other models of $f(X)$
straightforwardly.

Note that our model has a rescaling invariance. Suppose the solution
of $(\ref{Xsolli})$ is $X=X_0$, then we can always redefine
$\alpha\rightarrow \sqrt{X_0}\alpha$, $\beta\rightarrow
\beta/\sqrt{X_0}$, $c_1\rightarrow X_0 c_1$ and $c_2\rightarrow
X_0^2 c_2$, so that we get the solution $X=1$ after performing the
rescaling. So we set $X=1$ in the following calculation.

One can show that $\lambda/\Sigma$ and $c_s^2$ can be expressed as
\begin{equation}
  \frac{\lambda}{\Sigma}=\frac{2c_2}{c_1+6c_2}~,~~~c_s^2=1-2\frac{\lambda}{\Sigma}~.
\end{equation}
Combining with (\ref{Xsolli}), we can express the coefficients $c_1$
and $c_2$ as functions of $\lambda/\Sigma$ as
\begin{equation}
  c_1=-\frac{\frac{\alpha^2\beta^2}{3}\left(\frac{2}{-\frac{\lambda}{\Sigma}}+6\right)}
  {\frac{24}{\eta}+3+\left(\frac{6}{\eta}+1\right)\left(\frac{2}{-\frac{\lambda}{\Sigma}}\right)}~,~~~c_2=\frac{\frac{\alpha^2\beta^2}{3}}
  {\frac{24}{\eta}+3+\left(\frac{6}{\eta}+1\right)\left(\frac{2}{-\frac{\lambda}{\Sigma}}\right)}~.
\end{equation}
From the equation (\ref{fNL}), we see when $\lambda/\Sigma \ll -1$,
we can get a large and positive $f_{NL}^{equil}$. Note that this
case corresponds to $c_s\gg 1$. In order not to generate too large
tensor mode perturbation, we need $\epsilon \ll 1$, and mainly use
$\eta$ to generate a red spectrum for the scalar perturbations.

The parameter region $c_s\gg 1$ seems exotic, because this leads to a
superluminal propagation of the inflaton perturbations. However, as
discussed in \cite{Mukhanov:2005bu}, during inflation, the inflaton
field provides a time dependent background, which determines a
preferable coordinate frame. The superluminal propagation occurs only
in this special frame. So causality is not violated. This causality
issue is discussed in more detail in \cite{Babichev:2007dw}.

Comparing with  data, from $n_s\simeq 0.96$, and the observation
that $\epsilon$ is very small, we get $\eta\simeq 0.04$, so
$\beta\simeq 150\sqrt{2}$. If we assume $f_{NL}^{equil}\simeq 100$,
then we get $\lambda/\Sigma\simeq -810$, and $c_s\simeq 40$.
Finally, from the COBE normalization $P^{\zeta}\simeq 2.5\times
10^{-9}$, we get $\alpha^2/\epsilon\simeq 1.6\times 10^{-9}$.

Note that $\epsilon$ can still be chosen arbitrarily within the
experimental range. A different choice of $\epsilon$ leads to a
different tensor-to-scalar ratio $r$.

For example, when $r\simeq 0.3$, we have $\epsilon=4.7\times
10^{-4}$, $\alpha=8.7\times 10^{-7}$, $\phi=3.0\times 10^6$,
$c_1=-1.1\times 10^{-10}$, and $c_2=1.9\times 10^{-11}$.

At the first sight, these parameters seem to be rather unnatural.
While note that for simplicity, we have rescaled the parameters to
have $X=1$. If we rescale back the parameters such that
$-c_1/\cosh^2(\alpha\phi)\simeq 1$, then the parameters become
$\alpha=0.53$, $\beta=0.00034$, $\phi=4.8$, $X=2.6\times 10^{-12}$,
$c_1=-43$, and $c_2=2.7\times 10^{12}$. Note that $c_2$ still seems
to be too large in the Planck units. The largeness of $c_2$ implies
the existence of a new scale, for example, the string scale $M_s$.
If we recover the Planck mass, then
\begin{equation}\label{canof}
  f=-43X+\left(2.7\times
  10^{12}\times\frac{M_s^4}{M_{pl}^4}\right)\frac{X^2}{M_s^4}~.
\end{equation}
If the string scale is $M_s\sim 10^{-2}M_{pl}$ or $M_s\sim
10^{-3}M_{pl}$, then $c_2$ becomes of order 1. The above action
assumes the form of an effective action, with the mass scale $M_s$
playing the role of a physical cut-off.

As a matter of fact, the physical inflaton should be a ``nearly
canonical'' field
\begin{equation}\label{runphitd}
  \tilde{\phi}=\tilde{\phi}_0+\frac{2\sqrt{-c_1}}{\alpha}\arctan(e^{\alpha\phi}),
\end{equation}
with a first order canonical kinetic term
\begin{equation}
  \tilde{X}=-\frac{1}{2}g^{\mu\nu}\partial_{\mu}\tilde{\phi}\partial_{\nu}\tilde{\phi}=-\frac{c_1X}{\cosh^2(\alpha\phi)}
\end{equation}
when the Lagrangian is expanded. Here $\tilde{\phi}_0$ is a free
parameter, which can be fixed by hand. The normalization
$-c_1/\cosh^2(\alpha\phi)\simeq 1$ we have chosen facilitates our
discussion greatly. It implies $\tilde{X}=X$, hence the result
(\ref{canof}) still holds if we replace $X$ with $\tilde{X}$.
Numerically the inflaton $\tilde{\phi}=\tilde{\phi}_0+37$ in this
case.

As another example, when $r\simeq 10^{-3}$, we have
$\alpha=5.0\times 10^{-8}$, $\phi=1.1\times 10^8$, $c_1=-3.8\times
10^{-13}$, and $c_2=6.3\times 10^{-14}$. After rescaling to
$-c_1/\cosh^2(\alpha\phi)\simeq 1$, we have $\alpha=9.2$,
$\beta=1.2\times 10^{-6}$, $\phi=0.59$, $X=2.9\times 10^{-17}$,
$c_1=-1.3\times 10^{4}$, $c_2=7.3\times 10^{19}$ and
$\tilde{\phi}=\tilde{\phi}_0+39$.

Before proceeding to the next section, we discuss some physical
issues in the models we studied above. First we note that the
coefficient $c_1$ is negative, this is nice, since the function $g$
in equation (\ref{soluli}) is negative, thus the leading kinetic
term $X$ in energy is always positive and the relation
(\ref{runphitd}) is well-defined. However, the coefficient $c_2$ is
positive, this leads to a negative $X^2$ term in energy, and if $X$
is sufficiently large, this negative term will cause instability.
This problem can be eased by introducing a positive higher order
term in $X$.

We have introduced a mass scale $M_s$ above to indicate that these
models may be treated as an effective field theory arising in string
theory. Just as in the DBI action, high order terms in $X$ can be
regarded as stringy correction at the tree level. In an effective
action, operators with larger scaling dimensions are suppressed by
power of $1/M_s$. Here $X$ has dimension 4, so with each extra
factor $X$, a factor $1/M_s^4$ is introduced. As we have seen, these
high order terms are certainly important during inflation, as a
merely $X^2$ can help to produce a large $f_{NL}$. Nevertheless,
these terms become less important when the universe evolves to
regimes of low energy.

In the first model discussed above, the ``nearly canonical'' scalar
field assumes a value $\tilde{\phi}=\tilde{\phi}_0+37$ in the
reduced Planck unit. Unless we fine-tune $\tilde{\phi}_0$, this
value lies in the trans-Planckian regime. Since the potential
$V(\phi)$ is proportional to the square of the hyperbolic tangent
function, the scalar field $\phi$ rolls down towards smaller values
(the linear term $X$ dominates slightly over the quadratic term
$X^2$ in the kinetic energy). The second model still gives a
$\tilde{\phi}$ above the Planck scale without fine tuning, although
$\phi$ is smaller than the reduced Planck scale. Again, $\phi$ rolls
towards smaller values too. It is interesting to study carefully
whether $\tilde{\phi}>1$ will cause a disaster to our models,
following the arguments in \cite{Huang:2007gk,Huang:2007qz}.

\section{Reconstruction of Power-Law $k$-Inflation}\label{pl}
As has been discussed in \cite{ArmendarizPicon:1999rj}, the
Lagrangian of power-law $k$-inflation takes the form
\begin{equation}\label{ppl}
p(\phi,X)=\frac{1}{\phi^2}g(X)
\end{equation}
in general.

In  power-law $k$-inflation models, the equations of motion
(\ref{Friedmann}) are solved by
\begin{eqnarray}\label{solpl}
\nonumber \phi&=&\sqrt{2X}t,\\
\nonumber X&=&\frac{1}{2}\dot{\phi}^2=\mathrm{constant},\\
\nonumber a&\propto&t^{\sqrt{(2Xg_{,X}-g)/X}},\\
g_{,X}&=&\sqrt{\frac{2(2Xg_{,X}-g)}{3X}}.
\end{eqnarray}
While the slow-variation parameters in (\ref{slowvar}) and the
spectral indices are reduced to
\begin{eqnarray}\label{slowvarpl}
\nonumber &&\epsilon=\frac{3Xp_{,X}}{2Xp_{,X}-p},\\
\nonumber &&\eta=s=0,~~~~l=-2\epsilon,\\
&&n_s-1=-2\epsilon=n_T.
\end{eqnarray}
The equality of $n_s-1$ and $n_T$ in (\ref{slowvarpl}), and the
no-running condition
\begin{equation}
\alpha_s=\frac{dn_s}{d\ln k}=0
\end{equation}
are important features of power-law $k$-inflation. These features
can be used to test or rule out this class of models by future
experiments. Another feature is that the spectral indices depend on
$\epsilon$ exclusively, due to the vanishing of $\eta$ and $s$ in
this class of models.

As we have argued, a large positive $f_{NL}^{equil}$ in (\ref{fNL})
requires that $-\frac{\lambda}{\Sigma}\sim\mathcal{O}(10^3)$. On the
other hand, the experimentally favorite value
$n_s\simeq0.958\pm0.016$ indicates
$\epsilon\sim\mathcal{O}(10^{-2})$ by (\ref{slowvarpl}). So we can
see $-\frac{\epsilon\lambda}{\Sigma}\gtrsim1$. The value of $c_s$
can be experimentally determined by the aid of (\ref{r}).
Theoretically we cannot exclude the possibility that models might
exist with both $c_s^2\lesssim\epsilon$ and a large positive
$f_{NL}^{equil}$. But from now on we will concentrate on models with
$c_s^2\gg\epsilon$, which is much simpler. Using (\ref{fnlprecise}),
up to $\mathcal{O}(1)$, the equilateral non-Gaussianity estimator
becomes
\begin{eqnarray}\label{fNLpl}
f_{NL}^{equil}=-\frac{10}{81}\frac{\lambda}{\Sigma}-\frac{85}{324}\left(\frac{1}{c_s^2}-1\right)+\frac{10}{81}\left(29+4\gamma-84\ln\frac{3}{2}\right)\frac{\epsilon\lambda}{\Sigma}+\mathcal{O}(\frac{\epsilon}{c_s^2},\epsilon).
\end{eqnarray}

In this section we will reconstruct the power-law $k$-inflation with
$g(X)$ of some specific forms. The input parameters for
reconstruction are $\epsilon$, $c_s$ and $f_{NL}^{equil}$. The sound
speed $c_s$ may be translated by relation (\ref{r}) into
tensor-to-scalar ratio $r$, which is constrained by experiment. The
last line of (\ref{slowvarpl}) translates $\epsilon$ into $n_s$ or
$n_T$, so an experimental constraint on spectral indices is
equivalent to a constraint on $\epsilon$ in power-law $k$-inflation
models.

It is useful to note that
\begin{equation}
p_{,X}=\frac{1}{\phi^2}g_{,X},~~~~\frac{Xp_{,X}}{p}=\frac{Xg_{,X}}{g},
\end{equation}
\emph{etc.} in power-law $k$-inflation models (\ref{ppl}). The
expressions of $c_s^2$ and $\epsilon$ in (\ref{soundspd}) and
(\ref{slowvar}) can be rewritten in the form
\begin{eqnarray}\label{reeq12}
\nonumber \frac{Xg_{,X}}{g}&=&\frac{\epsilon}{2\epsilon-3}\equiv\xi_1,\\
\frac{X^2g_{,XX}}{Xg_{,X}}&=&\frac{1-c_s^2}{2c_s^2}\equiv\xi_2.
\end{eqnarray}
On the other hand, combining (\ref{soundspd}), (\ref{siglam}) and
(\ref{fNLpl}), we have
\begin{eqnarray}\label{reeq3}
\nonumber \frac{X^3g_{,XXX}}{X^2g_{,XX}}&=&-\frac{3}{2}+\frac{3}{1-c_s^2}\frac{\lambda}{\Sigma}\\
\nonumber &=&-\frac{3}{2}-3\left[1-\left(29+4\gamma-84\ln\frac{3}{2}\right)\epsilon\right]^{-1}\left[\frac{81f_{NL}^{equil}}{10(1-c_s^2)}+\frac{17}{8c_s^2}\right]\\
&\equiv&\xi_3.
\end{eqnarray}
Please note here
\begin{equation}
29+4\gamma-84\ln\frac{3}{2}\simeq-2.75
\end{equation}
numerically. Thanks to (\ref{reeq12}), the last equation of
(\ref{solpl}) may take a simple form as follows:
\begin{equation}\label{reeq4}
g_{,X}=\frac{2}{\epsilon}\equiv\xi_4.
\end{equation}
We have defined four parameters $\xi_1,\xi_2,\xi_3,\xi_4$ for later
use. In addition, it is necessary to check the condition
\begin{equation}\label{reeq5}
2Xg_{,X}-g>0
\end{equation}
dictated by $\rho>0$, otherwise the solution (\ref{solpl}) would
break down. After a few calculations, one can quickly confirm that
\begin{equation}\label{rhok}
\rho=\frac{1}{\phi^2}(2Xg_{,X}-g)=\frac{6X}{\phi^2\epsilon^2}=\frac{3}{\epsilon^2t^2}>0
\end{equation}
So the condition $\rho>0$ is always satisfied in power-law
$k$-inflation.

Equations (\ref{reeq12}-\ref{reeq4}) will be our main starting
point. We will reconstruct $g(X)$ of polynomial form
\begin{equation}\label{polyg}
g(X)=c_1X+c_2X^2+c_3X^3+c_4X^4
\end{equation}
in subsection \ref{polypl}, and that of DBI-like plus constant form
\begin{equation}\label{DBIcg}
g(X)=-(c_0+c_1X+c_2X^2)^{\frac{1}{2}}+c_3
\end{equation}
in subsection \ref{DBIcpl}. Subsection \ref{DBIpl} will concern the
DBI-like form
\begin{equation}\label{DBIg}
g(X)=-(c_0+c_1X+c_2X^2+c_3X^3)^{\frac{1}{2}}.
\end{equation}

\subsection{Power-Law Model \emph{I}: $p=\frac{1}{\phi^2}(c_1X+c_2X^2+c_3X^3+c_4X^4)$}\label{polypl}
For polynomial form (\ref{polyg}), the Lagrangian
\begin{equation}\label{polyp}
p=\frac{1}{\phi^2}(c_1X+c_2X^2+c_3X^3+c_4X^4).
\end{equation}
By substituting (\ref{polyg}) into (\ref{reeq12}-\ref{reeq4}), one
immediately gets
\begin{eqnarray}\label{polycoe}
\nonumber c_1&=&-\frac{\xi_4}{6\xi_1}(\xi_1\xi_2\xi_3-6\xi_1\xi_2+18\xi_1-24),\\
\nonumber c_2&=&\frac{\xi_4}{2X\xi_1}(\xi_1\xi_2\xi_3-5\xi_1\xi_2+12\xi_1-12),\\
\nonumber c_3&=&-\frac{\xi_4}{2X^2\xi_1}(\xi_1\xi_2\xi_3-4\xi_1\xi_2+8\xi_1-8),\\
c_4&=&\frac{\xi_4}{6X^3\xi_1}(\xi_1\xi_2\xi_3-3\xi_1\xi_2+6\xi_1-6).
\end{eqnarray}
During the reconstruction, $\xi_1,\xi_2,\xi_3,\xi_4$ may be traded
for $\epsilon$, $c_s$ and $f_{NL}^{equil}$ by
(\ref{reeq12}-\ref{reeq4}), while the value of $X$ is put by hand,
contingent on the scale of $\phi$ as will be shown in (\ref{polyX}).
Please note here $c_1$ is independent of $X$.

In principle, the Lagrangian (\ref{polyp}) can be reckoned as the
Lagrangian of a massless scalar with higher order corrections. To
see this, we have to redefine the scalar field as
\begin{equation}\label{polyphitd}
\tilde{\phi}=\sqrt{c_1}\ln\frac{\phi}{\phi_0},
\end{equation}
whose first order kinetic term
\begin{equation}\label{polyXtd}
\tilde{X}=-\frac{1}{2}g^{\mu\nu}\partial_{\mu}\tilde{\phi}\partial_{\nu}\tilde{\phi}=\frac{c_1X}{\phi^2}
\end{equation}
is canonical as we intend to show now. In terms of $\tilde{\phi}$
and $\tilde{X}$, the Lagrangian (\ref{polyp}) takes a ``nearly
canonical'' form
\begin{eqnarray}\label{canopolyp}
\nonumber p&=&\frac{1}{\phi^2}(c_1X+c_2X^2+c_3X^3+c_4X^4)\\
\nonumber &=&\tilde{X}+\frac{c_2}{c_1^2}\phi^2\tilde{X}^2+\frac{c_3}{c_1^3}\phi^4\tilde{X}^3+\frac{c_4}{c_1^4}\phi^6\tilde{X}^4\\
\nonumber &=&\tilde{X}+\tilde{c}_2\exp\left(\frac{2\tilde{\phi}}{\sqrt{c_1}}\right)\tilde{X}^2+\tilde{c}_3\exp\left(\frac{4\tilde{\phi}}{\sqrt{c_1}}\right)\tilde{X}^3+\tilde{c}_4\exp\left(\frac{6\tilde{\phi}}{\sqrt{c_1}}\right)\tilde{X}^4\\
&=&M_s^4\left(\frac{\tilde{X}}{M_s^4}+c_2^{*}\frac{\tilde{X}^2}{M_s^8}+c_3^{*}\frac{\tilde{X}^3}{M_s^{12}}+c_4^{*}\frac{\tilde{X}^4}{M_s^{16}}\right).
\end{eqnarray}

Using (\ref{Friedmann}), (\ref{sptrm}), (\ref{r}) and (\ref{solpl}),
(\ref{rhok}), one can confirm that
\begin{equation}\label{kspstm}
P^{\zeta}=\frac{4X}{r\pi^2\epsilon^2\phi^2}.
\end{equation}
Through this relation, given a normalization of power spectrum, the
scale of $\tilde{X}$ is dictated by
\begin{equation}\label{polyXtdscale}
\tilde{X}=\frac{c_1r\pi^2\epsilon^2}{4}P^{\zeta},
\end{equation}
while the scale of $X$ depends on $\phi_0$ and $\tilde{\phi}$ as
\begin{equation}\label{polyX}
X=\frac{r\pi^2\epsilon^2}{4}P^{\zeta}\phi_0^2\exp\left(\frac{2\tilde{\phi}}{\sqrt{c_1}}\right).
\end{equation}
In the rest of this subsection, we will set the ``canonical'' scalar
field $\tilde{\phi}\simeq0.01$, which is of the same order as
$\tilde{X}^{\frac{1}{4}}$ approximately. Since $\sqrt{c_1}\gg1$ in
(\ref{polyX}), our result will not change significantly if the scale
of $\tilde{\phi}$ is lowered down. That is to say, with
$\tilde{\phi}$ at a sub-Planckian energy scale, $X$ is determined by
$\phi_0$ but insensitive to $\tilde{\phi}$. We will not fix $X$ or
$\phi_0$ here, since they do not appear in our final result.

We choose parameters
\begin{equation}\label{input}
P^{\zeta}=2.5\times10^{-9},~~~~n_s\simeq0.97,~~~~f_{NL}^{equil}\simeq100,~~~~r\simeq0.1,
\end{equation}
and set
\begin{equation}
\tilde{\phi}\simeq0.01,~~~~\frac{M_s^4}{M_{pl}^4}\simeq5\times10^{-9},
\end{equation}
from (\ref{polycoe}), (\ref{polyphitd}), (\ref{polyXtdscale}) and
(\ref{polyX}), we can reconstruct the coefficients in Lagrangian
(\ref{canopolyp}) as
\begin{eqnarray}
\nonumber &&c_1\simeq4.6\times10^{4},~~~~c_2\simeq-2.1\times10^{18}\phi_0^{-2},\\
\nonumber &&c_3\simeq1.8\times10^{31}\phi_0^{-4},~~~~c_4\simeq-4.7\times10^{43}\phi_0^{-6},\\
\nonumber &&\tilde{c}_2\simeq-1.0\times10^{9},~~~~\tilde{c}_3\simeq1.9\times10^{17},~~~~\tilde{c}_4\simeq-1.1\times10^{25},\\
&&c_2^{*}\simeq-5.1,~~~~c_3^{*}\simeq4.8,~~~~c_4^{*}\simeq-1.3,
\end{eqnarray}
and
\begin{equation}
\phi\simeq\phi_0,~~~~X\simeq1.4\times10^{-13}\phi_0^2,~~~~\tilde{X}\simeq6.3\times10^{-9}M_{pl}^4\simeq1.3M_s^4.
\end{equation}

Note that selecting a value of $X$ is equivalent to choosing a
normalization of $\phi$. In the last line of (\ref{canopolyp}), we
have tuned $c_2^{*},c_3^{*},c_4^{*}$ to be of order unity by
choosing the string energy scale $M_s$. The first term in
(\ref{canopolyp}) recovers a canonical form apparently. One should
also note that contributions from each term in (\ref{canopolyp}) are
comparable. This is reasonable for the Lagrangian of a scalar field
with higher order corrections.

From (\ref{r}) and (\ref{slowvarpl}), one can show that for this set
of parameters, $\epsilon$ and $c_s^2$ take the values
\begin{equation}\label{ecs}
\epsilon\simeq0.015,~~~~c_s^2\simeq0.174.
\end{equation}
So we are sure that for the choice of parameters (\ref{input}), the
$\mathcal{O}(\frac{\epsilon}{c_s^2})$ contribution to
$f_{NL}^{equil}$ is indeed negligible in (\ref{fNLpl}).

\begin{figure}
\center{\includegraphics[width=0.45\textwidth]{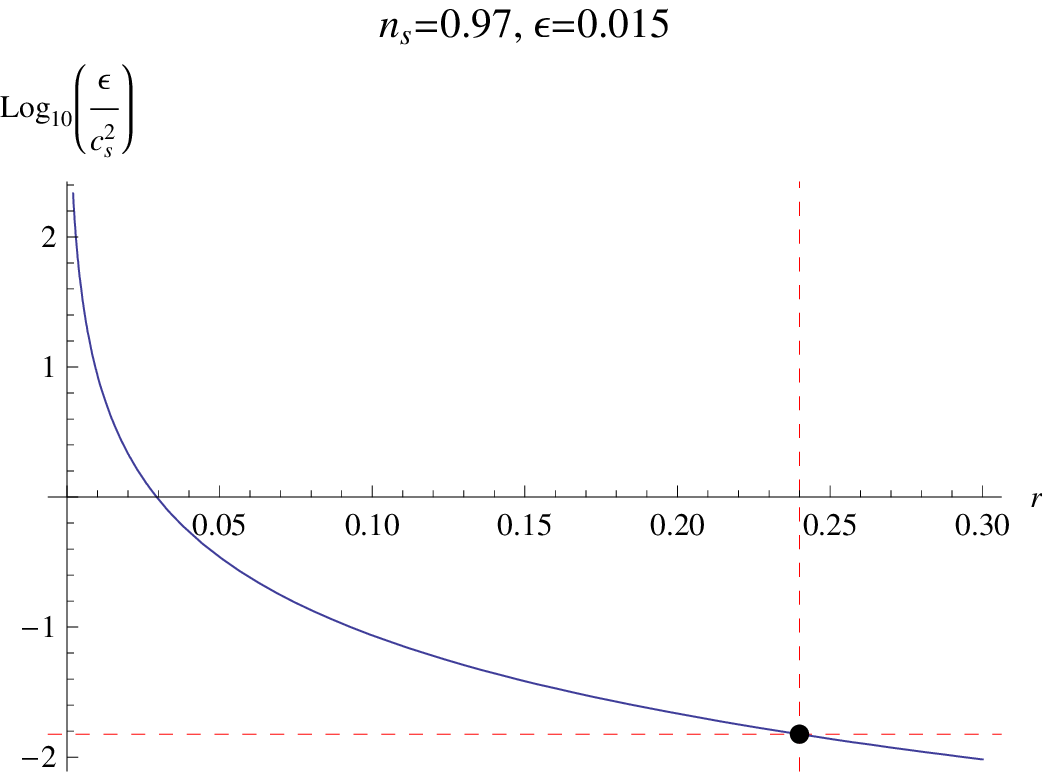}~~~~\includegraphics[width=0.45\textwidth]{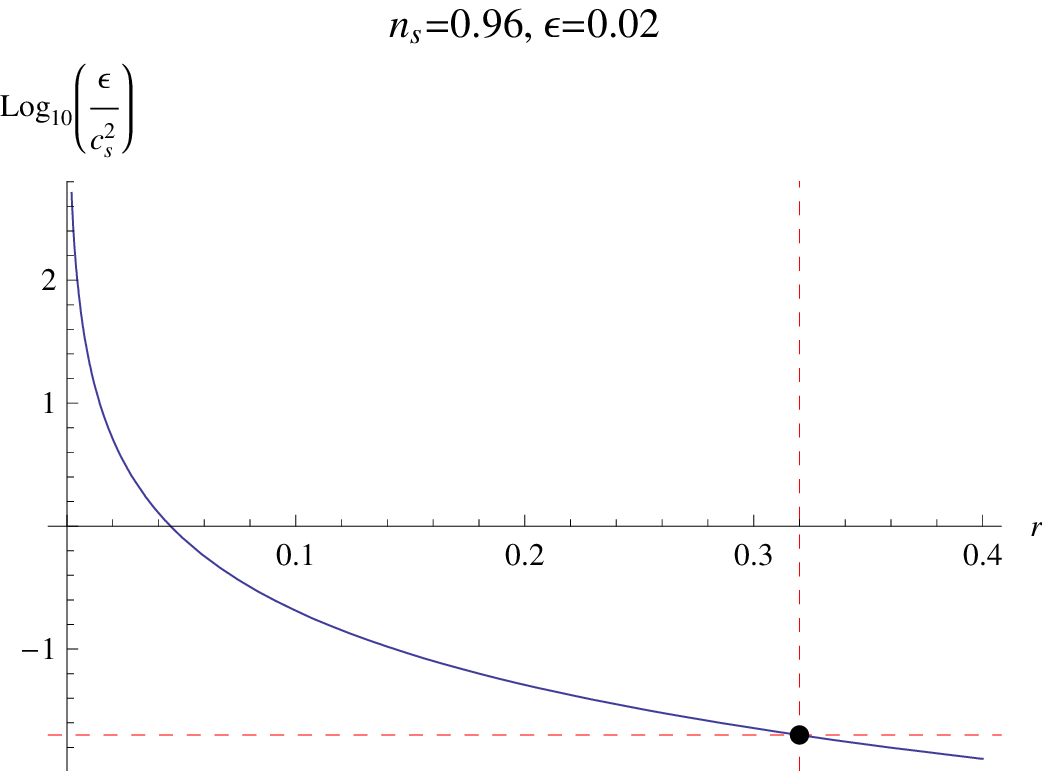}}\\
\caption{\textbf{The (logarithmic) ratio
$\log_{10}(\frac{\epsilon}{c_s^2})$ as a function of $r$. The solid
blue lines are plotted according to relation (\ref{ecsratio}). We
have set $n_s\simeq0.97$ in the left plot, and $n_s\simeq0.96$ in
the right one. This figure is valid for all of the power-law models
considered in section \ref{pl}. The dashed red lines are used to
highlight the black points corresponding to $c_s=1$, since we have
the additional constraint $c_s^2>1$ for power-law model \emph{II} in
subsection \ref{DBIcpl}.}}\label{ecsline}
\center{\includegraphics[width=0.5\textwidth]{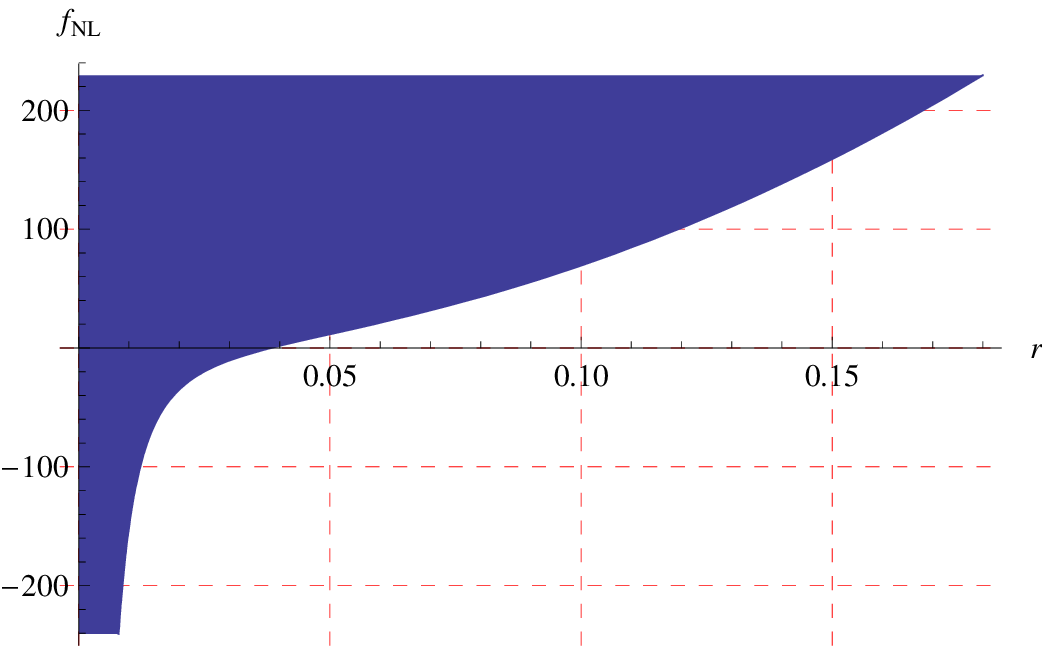}}\\
\caption{\textbf{The allowed $f_{NL}$-$r$ region (blue shaded) for
nearly canonical power-law model \emph{I}. This figure is plotted to
depict the condition (\ref{polycond}).}}\label{polycontour}
\end{figure}

As we have mentioned, by relations (\ref{r}) and (\ref{slowvarpl}),
the parameters $\epsilon$ and $c_s$ can be translated into $n_s$ and
$r$, which are experimentally constrained. The WMAP3 data alone
\cite{Spergel:2006hy} gives
\begin{equation}\label{wmap3}
n_s=0.958\pm0.016~\mathrm{(at~68\%~C.L.)},~~~~r_{0.002}<0.65~\mathrm{(at~95\%~C.L.)}.
\end{equation}
Therefore, more generally, given a reasonable value of $n_s$, we can
plot the ratio $\frac{\epsilon}{c_s^2}$ as a function of $r$,
\begin{equation}\label{ecsratio}
\frac{\epsilon}{c_s^2}=\frac{256\epsilon^3}{r^2}=\frac{32(1-n_s)^3}{r^2}.
\end{equation}
This has been done using solid blue lines in figure \ref{ecsline}.
The figure tells us that to neglect the
$\mathcal{O}(\frac{\epsilon}{c_s^2})$ contribution we should treat
our power-law models in large $r$ region (typically
$r\gtrsim\mathcal{O}(10^{-2})$), otherwise
$\mathcal{O}(\frac{\epsilon}{c_s^2})$ terms will be involved.

One should be aware that we choose $n_s=0.97$ instead of the
best-fit value $n_s=0.96$ in this section. This choice will be
essential in subsection 5.2, because models with $n_s=0.96$ and
$r<0.3$ can not be constructed in that subsection. In subsections
5.1 and 5.3, models with $n_s=0.96$ can nevertheless be constructed
as well.

If we assume $r\simeq0.3$ in (\ref{input}), we are not fortunate
enough to get a reasonable model here, because the reconstruction
gives values of $\phi$ with non-vanishing imaginary parts. This
indicates terms like $c_5X^5$ will be important for large $r$.

Let us see it in more details. In fact, the definition
(\ref{polyphitd}) implies $c_1>0$, namely
\begin{equation}\label{polycond}
f_{NL}^{equil}\gtrsim-0.059-\frac{0.015}{r^2}+7103r^2
\end{equation}
assuming the same choice of parameters $P^{\zeta}$ and $n_s$ as in
(\ref{input}). Clearly this condition is broken down when
$f_{NL}^{equil}\lesssim100$ and $r\simeq0.3$. The allowed region for
$f_{NL}$ as a function of $r$ is shaded in figure \ref{polycontour}
according to this condition.

For large $r$, although it is difficult to reconstruct the nearly
canonical Lagrangian (\ref{canopolyp}), such a model may be
available if we introduce a $c_5X^5$ term. In deed, when this term
is introduced, the condition (\ref{polycond}) will be relaxed, with
the form
\begin{equation}
f_{NL}^{equil}\gtrsim-0.059-\frac{0.015}{r^2}+7103r^2-2.5\times10^{-49}c_5\phi^8r^6,
\end{equation}
which is not hard to meet with reasonable values of parameters even
if $r$ is large.

Note that we introduced a mass scale $M_s$ for this class of models
again. All the coefficients look reasonable in this scale, and the
scalar field itself is in the sub-Planckian regime. This tells us
that this class of model as a candidate for explaining a large
$f_{NL}$ is attractive. Note also the highest term $X^4$ has a
negative coefficient, once again signaling instability. This may be
a general feature of all models accommodating a large positive
non-Gaussianity. Numerically, all the terms in $p$ are comparable in
magnitude.

\subsection{Power-Law Model \emph{II}: $p=\frac{1}{\phi^2}(-\sqrt{c_0+c_1X+c_2X^2}+c_3)$}\label{DBIcpl}
For the DBI-like plus constant form
\begin{equation}\label{DBIcp}
p=\frac{1}{\phi^2}(-\sqrt{c_0+c_1X+c_2X^2}+c_3),
\end{equation}
a similar computation gives
\begin{eqnarray}\label{DBIccoe}
\nonumber c_0&=&\frac{X^2\xi_4^2}{\xi_3^2}(\xi_3^2-3\xi_2\xi_3+6\xi_3+9),\\
\nonumber c_1&=&\frac{2X\xi_4^2}{\xi_3}(3\xi_2-\xi_3-3),\\
\nonumber c_2&=&\frac{\xi_4^2}{\xi_3}(\xi_3-3\xi_2),\\
c_3&=&\frac{X\xi_4}{\xi_3}\left(\frac{\xi_3}{\xi_1}+3\right),
\end{eqnarray}
and
\begin{equation}\label{DBIcbase}
\sqrt{c_0+c_1X+c_2X^2}=\frac{3\xi_4X}{\xi_3}=\frac{6X}{\epsilon\xi_3}.
\end{equation}
Since we concentrate on the case $\epsilon\geq0$, the result
(\ref{DBIcbase}) tells us that $\xi_3>0$. According to this
requirement and $f_{NL}^{equil}\gg1$, the definition of $\xi_3$ in
(\ref{reeq3}) suggests $c_s^2>1$ in this model. According to
(\ref{r}) and (\ref{slowvarpl}), this bound implies a constraint on
this model
\begin{equation}\label{csbound}
c_s=\frac{r}{8(1-n_s)}>1
\end{equation}
if $n_s<1$ and $f_{NL}^{equil}\gg1$.

Now let us try to recast the model into a more realistic form
\begin{eqnarray}\label{canoDBIcp}
\nonumber p&=&\frac{1}{\phi^2}(-\sqrt{c_0+c_1X+c_2X^2}+c_3)\\
\nonumber &=&-\frac{\sqrt{c_0}}{\phi^2}\sqrt{1-\frac{2}{\sqrt{c_0}}\phi^2\tilde{X}+\frac{4c_2}{c_1^2}\phi^4\tilde{X}^2}+\frac{c_3}{\phi^2}\\
\nonumber &=&-\tilde{c}_0\exp(-2\alpha\tilde{\phi})\sqrt{1-\frac{2}{\tilde{c}_0}\exp(2\alpha\tilde{\phi})\tilde{X}+\tilde{c}_2\exp(4\alpha\tilde{\phi})\tilde{X}^2}+\tilde{c}_3\exp(-2\alpha\tilde{\phi})\\
&=&M_s^4\left(-c_0^{*}\sqrt{1-\frac{2}{c_0^{*}}\frac{\tilde{X}}{M_s^4}+c_2^{*}\frac{\tilde{X}^2}{M_s^8}}+c_3^{*}\right),
\end{eqnarray}
which to the first order recovers a DBI action. This may be
accomplished by introducing
\begin{equation}\label{DBIphitd}
\tilde{\phi}=\sqrt{\frac{-c_1}{2\sqrt{c_0}}}\ln\frac{\phi}{\phi_0}=\frac{1}{\alpha}\ln\frac{\phi}{\phi_0},~~~~\alpha=\sqrt{\frac{2\sqrt{c_0}}{-c_1}},
\end{equation}
and correspondingly
\begin{equation}\label{DBIXtd}
\tilde{X}=-\frac{1}{2}g^{\mu\nu}\partial_{\mu}\tilde{\phi}\partial_{\nu}\tilde{\phi}=\frac{-c_1X}{2\sqrt{c_0}\phi^2}.
\end{equation}

Comparing (\ref{DBIphitd}), (\ref{DBIXtd}) with (\ref{kspstm}), one
quickly writes down
\begin{eqnarray}\label{DBIX}
\nonumber \tilde{X}&=&-\frac{c_1r\pi^2\epsilon^2}{8\sqrt{c_0}}P^{\zeta},\\
X&=&\frac{r\pi^2\epsilon^2}{4}P^{\zeta}\phi_0^2\exp(2\alpha\tilde{\phi}).
\end{eqnarray}
Again with a sub-Planckian inflaton $\tilde{\phi}$, we have
$\alpha\tilde{\phi}\ll1$ and thus $X$ is insensitive to the scale of
$\tilde{\phi}$. In the following we will assume the ``canonical''
scalar field and $\tilde{X}^{\frac{1}{4}}$ are roughly of the same
order, \emph{e.g.} $\tilde{\phi}\simeq10^{-3}$, but we will leave
$X$ and $\phi_0$ undetermined.

This time the condition $c_s^2>1$ put a tighter constraint on $r$,
see inequality (\ref{csbound}) and figure \ref{ecsline}. In each
plot of the figure, this condition is satisfied only for the solid
blue line below the black dot. So we choose $r\simeq0.3$. The other
input parameters of reconstruction are chosen to be the same as in
(\ref{input}). In addition we take
\begin{equation}
\tilde{\phi}\simeq10^{-3},~~~~\frac{M_s^4}{M_{pl}^4}\simeq5\times10^{-11}.
\end{equation}
With (\ref{DBIccoe}), (\ref{DBIphitd}) and (\ref{DBIX}) at hand, we
can perform a computation similar to the previous subsection,
resulted in coefficients
\begin{eqnarray}
\nonumber &&c_0\simeq3.1\times10^{-21}\phi_0^{4},~~~~c_1\simeq-1.5\times10^{-8}\phi_0^{2},\\
\nonumber &&c_2\simeq1.8\times10^{4},~~~~c_3\simeq-1.1\times10^{-8}\phi_0^{2},\\
\nonumber &&\tilde{c}_0\simeq5.6\times10^{-11},~~~~\tilde{c}_2\simeq3.2\times10^{20},~~~~\tilde{c}_3\simeq-1.1\times10^{-8},\\
&&c_0^{*}\simeq1.1,~~~~c_2^{*}\simeq0.81,~~~~c_3^{*}\simeq-220
\end{eqnarray}
in Lagrangian (\ref{canoDBIcp}), and other quantities
\begin{eqnarray}
\nonumber &&\phi\simeq\phi_0,~~~~X\simeq4.2\times10^{-13}\phi_0^2,~~~~\tilde{X}\simeq5.55\times10^{-11}M_{pl}^4\simeq1.1M_s^4,\\
&&\epsilon=0.015,~~~~c_s^2=1.56.
\end{eqnarray}
Hence this class of models can be considered as a DBI action with
higher order corrections.

Note that the coefficient $c_1$ has the desired sign, while the sign
of $c_2$ is positive and this leads to the kinetic energy unbounded
from below when $X$ is large. The last term $c_3/\phi^2$ dominates
the Lagrangian numerically, but other terms are important to drive
the inflation with appropriate spectral indices, otherwise we can
only have a de Sitter space.

\subsection{Power-Law Model \emph{III}: $p=-\frac{1}{\phi^2}\sqrt{c_0+c_1X+c_2X^2+c_3X^3}$}\label{DBIpl}
The Lagrangian for the DBI-like form (\ref{DBIg}) is
\begin{equation}\label{DBIp}
p=-\frac{1}{\phi^2}\sqrt{c_0+c_1X+c_2X^2+c_3X^3}.
\end{equation}
We still start from (\ref{reeq12}-\ref{reeq4}). After some
calculations, we obtain
\begin{eqnarray}\label{DBIcoe}
\nonumber c_0&=&X^2\xi_4^2(1-\xi_2)-\frac{X^2\xi_4^2}{3\xi_1^2}(\xi_1\xi_2\xi_3-3\xi_1\xi_2+6\xi_1-3),\\
\nonumber c_1&=&X\xi_4^2(3\xi_2-2)+\frac{X\xi_4^2}{\xi_1}(\xi_2\xi_3-2\xi_2+2),\\
\nonumber c_2&=&\xi_4^2(1-3\xi_2)+\frac{\xi_2\xi_4^2}{\xi_1}(1-\xi_3),\\
c_3&=&\frac{\xi_2\xi_4^2}{3X\xi_1}(3\xi_1+\xi_3),
\end{eqnarray}
and
\begin{equation}\label{DBIbase}
\sqrt{c_0+c_1X+c_2X^2+c_3X^3}=-\frac{3X\xi_4}{\xi_1}=\frac{2X(3-2\epsilon)}{\epsilon^2}.
\end{equation}

This model is all right if we do not demand the kinetic term be
canonical, with the value of $X$ or $\phi$ to be put by hand. For
instance, if we take $\phi\simeq0.1$ and (\ref{input}), by using
equations (\ref{kspstm}) and (\ref{DBIcoe}), we can get
\begin{eqnarray}\label{noncanoDBI}
\nonumber &&c_0\simeq-1.4\times10^{-20},~~~~c_1\simeq3.3\times10^{-5},~~~~c_2\simeq-2.4\times10^{10},~~~~c_3\simeq5.8\times10^{24},\\
&&X\simeq1.4\times10^{-15},~~~~\epsilon=0.015,~~~~c_s^2=0.174.
\end{eqnarray}
This naive model gives $f_{NL}^{equil}\simeq100$ as we desired.

Now turn to reconstruction of canonical models with Lagrangian
(\ref{DBIp}). Different from the previous subsection, for this class
of models, without introducing more terms, we cannot recover DBI
action to the first order. The key point is as follows.

Formally, employing (\ref{DBIphitd}-\ref{DBIX}), one may rewrite
Lagrangian (\ref{DBIp}) as
\begin{equation}\label{canoDBIp}
p=-\tilde{c}_0\exp(-2\alpha\tilde{\phi})\sqrt{1-\frac{2}{\tilde{c}_0}\exp(2\alpha\tilde{\phi})\tilde{X}+\tilde{c}_2\exp(4\alpha\tilde{\phi})\tilde{X}^2+\tilde{c}_3\exp(6\alpha\tilde{\phi})\tilde{X}^3}.
\end{equation}
However, the validity of (\ref{DBIphitd}) indicates $c_1<0$, namely
\begin{equation}\label{DBIcond}
f_{NL}^{equil}\lesssim0.11-\frac{0.015}{r^2}+5.6r^2
\end{equation}
when we make the same choice of parameters $P^{\zeta}$ and $n_s$ as
in (\ref{input}). Clearly it is impossible to arrange parameters
obeying the condition (\ref{DBIcond}) with $f_{NL}^{equil}\gg1$ and
$r<1$. In other words, even by tuning $M_s$ and the normalization of
$\phi$, it is impossible to rewrite (\ref{DBIp}) in the form of a
conventional DBI action with higher order corrections.

Nevertheless, we can still construct other models of the form
(\ref{DBIp}), whose leading order kinetic term is not canonical when
expanded. If we consider that class of models which are not ``nearly
canonical'', the model (\ref{noncanoDBI}) is able to reproduce the
desired $f_{NL}$.

If one prefers being restricted to ``nearly canonical'' models,
another $c_4X^4$ term will lend a hand, just as what happened in
subsection \ref{polypl}. Actually, when we take roughly
$c_4\simeq10^{36}\phi^{-4}$ and choose the assumption (\ref{input}),
a model
\begin{eqnarray}
\nonumber p&=&-\frac{1}{\phi^2}\sqrt{c_0+c_1X+c_2X^2+c_3X^3+c_4X^4}\\
\nonumber &=&-\tilde{c}_0e^{-2\alpha\tilde{\phi}}\sqrt{1-\frac{2}{\tilde{c}_0}e^{2\alpha\tilde{\phi}}\tilde{X}+\tilde{c}_2e^{4\alpha\tilde{\phi}}\tilde{X}^2+\tilde{c}_3e^{6\alpha\tilde{\phi}}\tilde{X}^3+\tilde{c}_4e^{8\alpha\tilde{\phi}}\tilde{X}^4}\\
&=&-c_0^{*}M_s^4\sqrt{1-\frac{2}{c_0^{*}}\frac{\tilde{X}}{M_s^4}+c_2^{*}\frac{\tilde{X}^2}{M_s^8}+c_3^{*}\frac{\tilde{X}^3}{M_s^{12}}+c_4^{*}\frac{\tilde{X}^4}{M_s^{16}}}
\end{eqnarray}
is constructed, with the main results
\begin{eqnarray}
\nonumber &&c_0\simeq2.3\times10^{-16}\phi_0^{4},~~~~c_1\simeq-7.4\times10^{-3}\phi_0^{2},\\
\nonumber &&c_2\simeq9.2\times10^{10},~~~~c_3\simeq-5.0\times10^{23}\phi_0^{-2},~~~~c_4\simeq10^{36}\phi_0^{-4},\\
\nonumber &&\tilde{c}_0\simeq1.5\times10^{-8},~~~~\tilde{c}_2\simeq6.75\times10^{15},~~~~\tilde{c}_3\simeq-1.5\times10^{23},~~~~\tilde{c}_4\simeq1.3\times10^{30},\\
&&c_0^{*}\simeq0.51,~~~~c_2^{*}\simeq6.1,~~~~c_3^{*}\simeq-4.1,~~~~c_4^{*}\simeq1.0
\end{eqnarray}
and
\begin{eqnarray}
\nonumber &&\tilde{\phi}\simeq0.01,~~~~\frac{M_s^4}{M_{pl}^4}\simeq3\times10^{-8},~~~~\alpha\simeq2.03\times10^{-3},\\
\nonumber &&\phi\simeq\phi_0,~~~~X\simeq1.4\times10^{-13}\phi_0^2,~~~~\tilde{X}\simeq3.4\times10^{-8}M_{pl}^4\simeq1.1M_s^4,\\
&&\epsilon=0.015,~~~~c_s^2=0.174.
\end{eqnarray}
This model is more complicated than (\ref{DBIp}), but it is
canonical when expanded in terms of $\tilde{X}$ to the first order.
What is more, it does not suffer from trans-Planckian effects,
because the inflaton $\tilde{\phi}$ is well below the Planck scale.

\section{Conclusion}\label{conclusion}
In this paper, we have constructed several models with a large
positive $f_{NL}$ in the WMAP convention
\cite{Komatsu:2001rj,Spergel:2006hy}. These models are general
single field inflation with higher order kinetic terms. In one class
of models, the inflation is driven by the potential term of the
inflaton, as shown in section \ref{run}. In section \ref{pl}, we
have given another class of models, in which inflation is driven by
non-conventional kinetic terms
\cite{ArmendarizPicon:1999rj,Garriga:1999vw}. In both classes of
models, due to the appearance of non-conventional kinetic terms, we
can arrange the parameters to produce $f_{NL}^{equil}\gg1$. These
are first examples to generate $f_{NL}^{equil}\gg1$ in general
single field $k$-inflation. They are different from ghost inflation
\cite{ArkaniHamed:2003uz} since we restricted our discussion in
$k$-inflation and did not introduce terms like $\nabla^2\phi$ in our
Lagrangian.

The common features of these models are as following:
\begin{enumerate}
\item Typically
we need to introduce four parameters in a model, although there are
only three data to fit, the COBE normalization of the two point
function, the spectral index $n_s$ and the non-Gaussianity parameter
$f_{NL}$. Introduction of four parameters is not strictly necessary.
For example, for models studied in section 4, we introduced two
parameters $\alpha$ and $\beta$ in order to have a model easy to
solve. In section 5, we introduced four parameters $c_i$ since we
try to also fit a parameter $r$ whose value is not fixed by
experiments yet. However, a upper bound on $r$ is set, so it is
necessary to make sure that our model does not violate this bound.
\item High order terms in $X$ are absolutely necessary, and luckily
these terms can be viewed as operators of high dimensions in an
effective field theory with a mass cut-off lower than the Planck
scale. The bad news is that, the highest order term in general
triggers an instability if we do not introduce more high order
terms.
\end{enumerate}

\section*{Acknowledgments}

We thank Qing-Guo Huang for discussions. We are grateful to KITPC
for a fruitful workshop on string cosmology, during which this work
was initiated. We thank all the speakers on the topic of
non-Gaussianity for the education, especially Komatsu for his
informative talk. We thank Henry Tye especially, for his emphasis on
the importance of the non-Gaussianity in discriminating among
inflation models. This work was supported by grants from NSFC, a
grant from Chinese Academy of Sciences and a grant from USTC.

\appendix

\section{Calculation of $R^{equil}(k)$}\label{CalRk}
In the equilateral triangle case, the function $R(k_1,k_2,k_3)$
defined in \cite{Chen:2006nt} takes the form
\begin{eqnarray}
\nonumber R^{equil}(k)&=&R(k_1,k_2,k_3)|_{k_1=k_2=k_3=k}\\
&=&3k^3\mathrm{Re}\left[\int_{0}^{\infty}(1-2ix)e^{-2ix}h^{*}(x)dx\right]
\end{eqnarray}
with
\begin{equation}
h(x)=-2ie^{ix}+ie^{-ix}(1+ix)[\mathrm{Ci}(2x)+i\mathrm{Si}(2x)]-i\pi\sin x+i\pi x\cos x.
\end{equation}
With the help of the relation
\begin{eqnarray}
\nonumber \mathrm{Ci}(2x)-i\mathrm{Si}(2x)&=&-\int_{2x}^{\infty}\frac{\cos tdt}{t}-i\int_{0}^{2x}\frac{\sin tdt}{t}\\
&=&-\int_{2x}^{\infty}\frac{e^{-it}dt}{t}-\frac{i\pi}{2}
\end{eqnarray}
and the equality
\begin{equation}
\int_{-2ix}^{\infty}\frac{e^{-it}dt}{t}=\int_{2x}^{\infty}\frac{e^{-u}du}{u}~~~~\mathrm{for}~x\geq0,
\end{equation}
after an analytical continuation $x\rightarrow-ix$, we can transform
it into
\begin{eqnarray}
\nonumber R^{equil}(k)&=&3k^3\mathrm{Re}\left\{-i\int_{0}^{\infty}(1-2x)e^{-2x}\right.\\
\nonumber &&\times\left.\left[\left(2i-\frac{\pi}{2}(1+x)\right)e^{-x}+i(1-x)e^{x}\left(\int_{-2ix}^{\infty}\frac{e^{-it}dt}{t}\right)\right]dx\right\}\\
\nonumber &=&3k^3\int_{0}^{\infty}2(1-2x)e^{-3x}dx\\
&&+3k^3\int_{0}^{\infty}(1-x)(1-2x)e^{-x}\left(\int_{2x}^{\infty}\frac{e^{-u}du}{u}\right)dx.
\end{eqnarray}
The first term can be easily evaluated with a result
$\frac{2}{3}k^3$. To calculate the second term, let us note that
\begin{equation}
\int_{0}^{\infty}...dx=\lim_{\epsilon\rightarrow0^{+}}\int_{\epsilon}^{\infty}...dx
\end{equation}
and consider the integral
\begin{eqnarray}
\nonumber &&\lim_{\epsilon\rightarrow0^{+}}\left[\int_{\epsilon}^{\infty}(1-x)(1-2x)e^{-x}\left(\int_{2x}^{\infty}\frac{e^{-u}du}{u}\right)dx\right]\\
\nonumber &=&\lim_{\epsilon\rightarrow0^{+}}\left[(2+\epsilon+2\epsilon^2)e^{-\epsilon}\int_{2\epsilon}^{\infty}\frac{e^{-u}du}{u}-2\int_{\epsilon}^{\infty}\frac{e^{-3x}dx}{x}-\left(\frac{5}{9}+\frac{2}{3}\epsilon\right)e^{-3\epsilon}\right]\\
\nonumber &=&\lim_{\epsilon\rightarrow0^{+}}\left(2\int_{2\epsilon}^{\infty}\frac{e^{-u}du}{u}-2\int_{\epsilon}^{\infty}\frac{e^{-3x}dx}{x}-\frac{5}{9}\right)\\
\nonumber &=&-\frac{5}{9}+\lim_{\epsilon\rightarrow0^{+}}\left(2\int_{2\epsilon}^{\infty}\frac{e^{-u}du}{u}-2\int_{3\epsilon}^{\infty}\frac{e^{-u}du}{u}\right)\\
\nonumber &=&-\frac{5}{9}+\lim_{\epsilon\rightarrow0^{+}}2\int_{2\epsilon}^{3\epsilon}\frac{(1-u)du}{u}\\
&=&-\frac{5}{9}+2\ln\frac{3}{2}.
\end{eqnarray}
Here we have integrated by parts and used the equalities
\begin{equation}
\nonumber \frac{d}{dx}\int_{2x}^{\infty}\frac{e^{-u}du}{u}=-\frac{e^{-2x}}{x}.
\end{equation}
As the last result,
\begin{equation}
R^{equil}(k)=\frac{2}{3}k^3+3k^3\left(-\frac{5}{9}+2\ln\frac{3}{2}\right)=\left(-1+6\ln\frac{3}{2}\right)k^3
\end{equation}

\end{document}